\documentclass[conference]{IEEEtran}

\usepackage{cite}
\usepackage{algorithmic}
\usepackage{array}
\usepackage{graphicx,cite,epsfig,amssymb,amsmath}

\begin{document}
%
\title{A Non-Cooperative Method for Path Loss Estimation in Femtocell Networks}

\author{\IEEEauthorblockN{Qinliang Su\IEEEauthorrefmark{1}, Aiping
Huang\IEEEauthorrefmark{1}\IEEEauthorrefmark{2}, Zhaoyang
Zhang\IEEEauthorrefmark{1}\IEEEauthorrefmark{2}, Kai
Xu\IEEEauthorrefmark{3}, and Jin Yang\IEEEauthorrefmark{3}}

\IEEEauthorblockA{\IEEEauthorrefmark{1} Institute of Information and
Communication Engineering, Zhejiang University Hangzhou 310027,
China} \IEEEauthorblockA{\IEEEauthorrefmark{2} Zhejiang Provincial
Key Laboratory of Information Network Technology, Hangzhou 310027,
China}\IEEEauthorblockA{\IEEEauthorrefmark{3} China Broadband
Communication Research Lab, Applied Research Center, Motorola}
E-mail: aiping.huang@zju.edu.cn, kaixu@motorola.com}


%


\maketitle

\begin{abstract}
A macrocell superposed by indoor deployed femtocells forms a
geography-overlapped and spectrum-shared two-tier network, which can
efficiently improve coverage and enhance system capacity. To reduce
inter-tier co-channel interference, femtocell user should choose
suitable access channel according to the path losses between itself
and the macrocell users. Path loss should be estimated
non-cooperatively since information exchange is difficult between
macrocell and femtocells. In this paper, a novel method is proposed
for femtocell user to estimate the path loss between itself and any
macrocell user independently. According to the adaptive modulation
and coding (AMC) mode information broadcasted by the macrocell base
station (BS), femtocell user first estimates the path loss between
BS and a macrocell user by using Maximum a Posteriori (MAP) method.
The probability density function (PDF) and statistics of the
transmission power of the macrocell user are then derived. According
to the sequence of received power values from the macrocell user,
femtocell user estimates the path loss between itself and the
macrocell user by using minimum mean square error (MMSE) method.
Simulation results show that the proposed method can efficiently
estimate the path loss between any femtocell user and any macrocell
user in all kinds of conditions.
\end{abstract}


%
\IEEEpeerreviewmaketitle

\section{Introduction}
The two-tier network formed by deploying femtocells in the coverage
of an already existed macrocell can improve indoor coverage and
enhance system overall capacity by sharing spectrum between these
two tiers \cite{c1}. In the view of cognitive radio \cite{Haykin05},
the macrocell is the primary tier whose users and base station (BS)
communicate to each other and have a priority to use channels, the
femtocells are of the secondary tier whose users can communicate
with their access points (APs) under the premise of no obvious
influence on the communication in primary tier network. For the
conciseness of presentation, femtocell user is called SU (secondary
user), while macrocell user is called PU (primary user) in this
article.

The spectrum sharing in this two-tier network leads to not only high
spectrum efficiency but also serious inter-tier interference. If SU
knows the path losses between itself and all the PUs, then each SU
can choose the channel occupied by a PU far away from it so as to
reduce the inter-tier co-channel interference. Therefore, the
estimation of path loss between SU and PU plays a very important
role in femtocell networks.

Existing methods for estimating the path loss between two locations
can be classified into two types, that is, model-based method and
measurement-based method. In the model-based methods \cite{c3}, the
estimating end calculates the path loss according to the known
propagation loss model and distance between the two ends. However,
for the SU, which works as the estimating end, it is very difficult
to obtain the distances between itself and a PU in a real two-tier
network. On the other hand, in the measurement-based methods
\cite{c4,c5,c6}, the estimating end first obtains the transmission
power of the other end, and then calculates the path loss as the
ratio of the received power to the transmission power. Here, the
transmission power value is either fixed and known to the estimating
end, such as the pilot signal, or is included in a packet which is
transmitted to the estimating end. In the literature, existing
measurement-based methods require the estimating end to be informed
explicitly of the transmission power of the other end. That is, the
two ends should work cooperatively. However, PU's transmission power
depends on its location, channel status, required SINR etc, and
inter-tier information exchange should not be assumed in a real
two-tier network. Therefore, for an SU (the estimating end), it is
very hard to know the value of a PU's transmission power.

To solve the above problem, this paper proposes a novel path loss
estimation method for an SU to independently estimate the path loss
between itself and PUs in two-tier network. Similar to the existing
measurement-based methods, the proposed method also consists of two
steps. Firstly, SU demodulates the BS broadcast information and
obtains the downlink and uplink adaptive modulation and coding (AMC)
modes used by the transmitting signals of BS and PU. SU estimates
the path loss between the PU and BS according to the AMC modes, and
derives the probability density function (PDF) and statistics of the
PU's transmission power. Secondly, after measuring the received
power from the PU, SU estimates the path loss between itself and the
PU by exploiting the relation among transmission power, received
power and path loss. This method only utilizes the AMC mode
information broadcast by BS and the statics of channel between BS
and PU, while does not need any information exchange between SU and
PU. This non-cooperative estimation method does not have any impact
on the macrocell network.

This paper is organized as follows. The mathematical models of AMC
mode assigned by BS for PU transmission signal, PU transmission
power and SU received power are established in Section II. In
Section III, the PDF and statistics of PU transmission power are
derived. The estimation method of path loss between SU and PU is
given in Section IV. Simulation results are given in Section V,
followed by conclusions in Section VI.

\section{System model}
The considered system model is shown in Fig. 1. The macrocell is a
normal cell of a cellular network, in which PU and BS communicate to
each other and have priority to use the channels. BS broadcasts
information of AMC modes of the transmission signals in the downlink
and uplink for demodulation and modulation in the cell. By sharing
spectrum with PU, SU can communicate with its access point (AP)
under the condition that the macrocell user is not affected
obviously.
\begin{figure}[!t]
\centering
\includegraphics[width=0.4\textwidth]{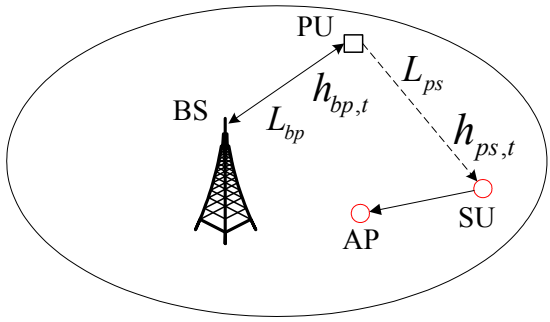}
\caption{System structure} \label{system_structure}
\end{figure}
\subsection{AMC mode assigned by BS}
In the main stream cellular networks, such as WiMAX and LTE, BS uses
a constant transmission power in the downlink for convenience, and
assigns an appropriate AMC mode for BS transmission signal according
to channel status so as to make the best use of channel capacity.
Denote the transmission power of BS as $P_0$. Then, the SINR of the
received signal of PU at time instant $i$ can be written as
\begin{equation}
{S_{d,i}  = \frac{{P_0 \left| {h_{bp,i} } \right|^2 }}{{\sigma ^2
L_{bp} }},\;\;\;i = 1,\;2,\; \cdots ,\;I},
\end{equation}
where the subscript $d$ represents downlink; $h_{bp,i}$ is the
channel response between BS and PU at time instant $i$, whose
absolute value obeys the Rayleigh distribution; $\sigma ^2$ is the
average power of background noise; $L_{bp}$ is the downlink path
loss between BS and PU; $I$ is the number of observations.

Suppose there are $M$ available AMC modes, where mode sequence
number $m = 1,2, \cdots ,M$. The order of the $(m+1)$-th mode is
higher than that of the $m$-th mode, and the required receiving SINR
of the $(m+1)$-th mode is also higher. The sequence number of AMC
mode assigned by BS for its transmission signal at time instant $i$
can be represented as
\begin{equation}
{m_{d,i}  = \Psi \left( {S_{d,i} } \right),\;\;\;i = 1,\;2,\; \cdots
,\;I},
\end{equation}
where $m_{d,i}  \in \{ 1,2,...M\}$; $\Psi \left(  \cdot  \right)$
stands for an assignment function, which produces a sequence number
of AMC mode which satisfies the required receiving SINR requirement
and maximizes the transmission rates. The sequence number of AMC
modes assigned by BS for its transmission signals in the last $I$
instants composes a vector ${\bf{m}}_d = [m_{d,1} ,\; \cdots
,m_{d,i} ,\; \cdots ,\;m_{d,I} ]^{\rm{T}}$.
\subsection{PU transmission power}
In the uplink, PU adjusts its transmission power so that the
received SINR of BS satisfies the requirement of adopted AMC mode
exactly, as applied in WiMAX and LTE systems. The transmission power
of PU at time instant $i$ can be written as
\begin{equation}
{P_{t,i}  = \left\{ \begin{array}{l}
 P_{\min } ,\;\;\;\;\;\;\;\;\;\;\;\;\tilde P_{t,i}  < P_{\min }  \\
 \tilde P_{t,i} ,\;\;\;\;\;P_{\min }  \le \tilde P_{t,i}  \le P_{\max }  \\
 P_{\max } ,\;\;\;\;\;\;\;\;\;\;\;\;\tilde P_{t,i}  > P_{\max }  \\
 \end{array} \right.,\;\;\;\;i = 1,\;2,\; \cdots ,\;I},
\end{equation}
where the subscript $t$ represents transmission; $P_{\min}$ and
$P_{\max}$ are the minimum and maximum allowed transmission powers
of PU. When no power restriction is imposed, in order to satisfy the
required receiving SINR, the PU's transmission power can be
expressed as
\begin{equation}
{\tilde P_{t,i}  = \frac{{\Omega \left( {m_{u,i} } \right)L_{pb}
}}{{\left| {h_{pb,i} } \right|^2 }}\sigma ^2},
\end{equation}
where $m_{u,i}$ is the sequence number of the AMC mode assigned by
BS for PU's transmission signal at time instant $i$; subscript $u$
represents uplink; $\Omega \left( m \right)$ stands for a mapping
function, which produces the required minimum SINR according to the
sequence number of the AMC mode $m$; $L_{pb}$ is the path loss
between PU and BS in the uplink; $h_{pb,i}$ is the channel response
between PU and BS at time instant $i$.
\subsection{Received power of SU}
When PU transmits signal to BS with power $P_{t,i}$ at time instant
$i$, the SU's received power of the signal transmitted by PU can be
expressed as
\begin{equation}
{\begin{array}{l}
 P_{r,i}  = \frac{{P_{t,i} \left| {h_{ps,i} } \right|^2 }}{{L_{ps} }} + P_{N,i}  \\
 \;\;\;\;\; = P_{t,i} \left| {h_{ps,i} } \right|^2 x + P_{N,i} ,\;\;\;\;i = 1,\;2,\; \cdots ,\;I \\
 \end{array}},
\end{equation}
where subscript $r$ represents receiving; $h_{ps,i}$  is the channel
response between PU and SU at time instant $i$; $x$ is the inversion
of the path loss between PU and SU; $P_{N,i}$ is the power of
background noise at time instant $i$, which is a random variable
following the $\chi ^2$ distribution of one degree of freedom with
mean value equal to $\sigma ^2$ . Equation $(5)$ can be expressed in
matrix form as
\begin{equation}
{{\bf{P}}_r  = diag\left( {{\bf{P}}_t } \right) \cdot {\bf{h}}_{ps}
x + {\bf{P}}_N},
\end{equation}
where ${\bf{P}}_r$ is the received power vector, and ${\bf{P}}_r  =
[P_{r,1} ,\; \cdots ,P_{r,i} ,\; \cdots ,\;P_{r,I} ]^{\rm{T}}$;
${\bf{P}}_t$ is the transmission power vector, and ${\bf{P}}_t  =
[P_{t,1} ,\; \cdots ,P_{t,i} ,\; \cdots ,\;P_{t,I} ]^{\rm{T}}$;
$diag\left( {{\bf{P}}_t } \right)$ is a diagonal matrix composed of
the elements in vector ${\bf{P}}_t$; ${\bf{h}}_{ps}$ is the channel
response vector, and ${\bf{h}}_{ps}  = [\left| {h_{ps,1} } \right|^2
, \cdots ,\;\left| {h_{ps,i} } \right|^2 ,\; \cdots ,\;\left|
{h_{ps,I} } \right|^2 ]^{\rm{T}}$; ${\bf{P}}_N$ is the noise power
vector, and $ {\bf{P}}_N = [P_{N,1} ,\; \cdots ,P_{N,i} ,\; \cdots
,\;P_{N,I} ]^{\rm{T}}$.
\section{PDF and statistics of PU transmission power}
In order to estimate the path loss between SU and PU, SU needs to
know the transmission power of PU. In this section, SU first
estimates the path loss between PU and BS by exploiting the
information of AMC mode assigned by BS in the downlink and then
derives the PDF and statistics of PU transmission power
${\bf{P}}_t$.
\subsection{Estimation of path loss $L_{pb}$ between PU and BS}
We first derive the joint PDF of the AMC mode sequence number vector
${\bf{m}}_d$ and the path loss $L_{bp}$ and then estimate the path
loss $L_{pb}$ by using MAP method.

According to Equations $(1)$ and $(2)$, when $L_{bp}$ is known, the
probability that BS assigns the $m_{d,i}$-th AMC mode for its
transmission signal at time instant $i$ can be expressed as
\begin{equation}
{\begin{array}{l}
 P\left( {m_{d,i} |L_{bp} } \right) = P\left( {\Omega \left( {m_{d,i} } \right) \le S_{d,i}  < \Omega \left( {m_{d,i}  + 1} \right)} \right) \\
 \;\;\;\;\;\;\;\;\;\;\;\;\;\; = P\left( {\frac{{\Omega \left( {m_{d,i} } \right)L_{bp} \sigma ^2 }}{{P_0 }} \le \left| {h_{bp,i} } \right|^2  < \frac{{\Omega \left( {m_{d,i}  + 1} \right)L_{bp} \sigma ^2 }}{{P_0 }}} \right) \\
 \;\;\;\;\;\;\;\;\;\;\;\;\;\; = F_y \left( {\frac{{\Omega \left( {m_{d,i}  + 1} \right)L_{bp} \sigma ^2 }}{{P_0 }}} \right) - F_y \left( {\frac{{\Omega \left( {m_{d,i} } \right)L_{bp} \sigma ^2 }}{{P_0 }}} \right) \\
 \;\;\;\;\;\;\;\;\;\;\;\;\;\; = e^{ - \;\frac{{\Omega \left( {m_{d,i} } \right)L_{bp} \sigma ^2 }}{{P_0 }}}  - e^{ - \;\frac{{\Omega \left( {m_{d,i}  + 1} \right)L_{bp} \sigma ^2 }}{{P_0 }}}  \\
 \end{array}},
\end{equation}
where the third equality is because $\left| {h_{bp,i} }
\right|^2$ is a random variable following the standard $\chi ^2$
distribution of two degrees of freedom , whose cumulative
distribution function (CDF) is $F_y (y) = 1 - e^{ - y}$ \cite{c7}.

Suppose PU is uniformly distributed in the Macrocell. Then, the PDF
of the distance between PU and BS, denoted as $z$, is
\begin{equation}
{f_z (z) = \frac{{2z}}{{R_0^2  - R_{\min }^2 }}},
\end{equation}
where $R_0$ is the radius of the Macrocell, and $R_{min}$ is the
allowed minimum distance between PU and BS. Shadowing effect does
not need to be considered since its time scale is usually much
larger than the observation period. Without considering shadowing,
we have $L_{bp} = L_0 z^\alpha$, where $L_0$ is the path loss at
unit distance, and $\alpha$ is the attenuation factor. Applying the
relation into Equation $(8)$, we obtain the PDF of $L_{bp}$ as
\begin{equation}
{\begin{array}{l}
 f_L (L_{bp} ) = \left. {f_z \left( z \right) \cdot \frac{{dz}}{{dL_{bp} }}} \right|_{z = \left( {\frac{{L_{bp} }}{{L_0 }}} \right)^{\frac{1}{\alpha }} }  \\
 \;\;\;\;\;\;\;\;\;\;\; = \frac{2}{{\alpha L_0 (R_0^2  - R_{\min }^2 )}}\left( {\frac{{L_{bp} }}{{L_0 }}} \right)^{\frac{2}{\alpha }\; - \;1}  \\
 \end{array}}.
\end{equation}
According to Equations $(7)$ and $(9)$, and under the assumption of
i.i.d Rayleigh fading channel, the joint PDF of ${\bf{m}}_d$ and
$L_{bp}$ is
\begin{equation}
{\begin{array}{l}
 f_{{\bf{m}},L} ({\bf{m}}_d ,\;L_{bp} ) = P({\bf{m}}_d |L_{bp} ) \cdot f_L (L_{bp} ) \\
 \;\;\;\;\;\;\;\;\;\;\;\;\;\;\;\;\;\;\;\; = P(m_{d,1} |L_{bp} ) \cdots P(m_{d,I} |L_{bp} ) \cdot f_L (L_{bp} ) \\
 \;\;\;\;\;\;\;\;\;\;\;\;\;\;\;\;\;\;\;\; = \frac{2}{{\alpha L_0 (R_0^2  - R_{\min }^2 )}}\left( {\frac{{L_{bp} }}{{L_0 }}} \right)^{\frac{2}{\alpha }\; - \;1}  \\
 \;\;\;\;\;\;\;\;\;\;\;\;\;\;\;\;\;\;\;\;\;\; \cdot \prod\limits_{i = 1}^T {\left( {e^{ - \;\frac{{\Omega \left( {m_{d,i} } \right)L_{bp} \sigma ^2 }}{{P_0 }}}  - e^{ - \;\frac{{\Omega \left( {m_{d,i}  + 1} \right)L_{bp} \sigma ^2 }}{{P_0 }}} } \right)}  \\
 \end{array}},
\end{equation}
where the second equality is held because the variables $m_{d,i}
\;(i = 1,\;2,\; \cdots ,\;I)$ are independent when $L_{bp}$ is
known.

After obtaining the AMC mode sequence number vector ${\bf{m}}_d$ by
demodulating and recording $m_{d,i} \;(i = 1,\;2,\; \cdots ,\;I)$,
SU can estimate the downlink path loss between BS and PU by using
MAP method as follows
\begin{equation}
{\hat L_{bp}  = \mathop {\max }\limits_{L_{bp} } \;f_{{\bf{m}},L}
({\bf{m}}_d ,\;L_{bp} )}.
\end{equation}

The estimate of uplink path loss between PU and BS, $\hat L_{pb}$,
is equal to $\hat L_{bp}$ if time division duplex (TDD) mode is
adopted, while $\hat L_{pb}$ equals $\hat L_{bp}$ plus a constant if
frequency division duplex (FDD) mode is adopted. The constant can be
derived from theoretical model of propagation loss, by substituting
the carrier frequency difference between uplink and downlink into
the model.
\subsection{PDF and statistics of PU transmission power ${\bf{P}}_t$}
From Equations $(3)$ and $(4)$, it can be seen that PU's
transmission power at time instant $i$, $P_{t,i}$, only depends on
the channel response $h_{pb,i}$ when $L_{pb}$ and $m_{u,i}$ are
known. Since variable $\left| {h_{pb,i} } \right|^2$ follows the
standard $ \chi ^2$ distribution of two degrees of freedom, the PDF
of $ P_{t,i}$ can be derived as
\begin{equation}
{f_P \left( {P_{t,i} } \right) = \left\{ \begin{array}{l}
 e^{ - \frac{{\Omega \left( {m_{u,i} } \right)L_{pb} \sigma ^2 }}{{P_{\min } }}} \delta \left( {P_{t,i}  - P_{\min } } \right)\;\;\;\;P_{t,i}  = P_{\min }  \\
 \frac{{\Omega \left( {m_{u,i} } \right)L_{pb} \sigma ^2 }}{{\left( {P_{t,i} } \right)^2 }}e^{ - \frac{{\Omega \left( {m_{u,i} } \right)L_{pb} \sigma ^2 }}{{P_{t,i} }}} ,\;P_{\min }  < P_{t,i}  < P_{\max }  \\
 1 - e^{ - \frac{{\Omega \left( {m_{u,i} } \right)L_{pb} \sigma ^2 }}{{P_{\max } }}} \delta \left( {P_{t,i}  - P_{\max } } \right),\;P_{t,i}  = P_{\max }  \\
 \end{array} \right.}.
\end{equation}
According to Equation $(12)$, the mean value $\bar P_{t,i}$ and mean
square value $\overline {P_{t,i}^2 }$ of PU's transmission power at
instant $i$ can be written as
\begin{equation}
{\bar P_{t,i}  = \int_{P_{\min } }^{P_{\max } } {P_{t,i} f_P \left(
{P_{t,i} } \right)} dP_{t,i}},
\end{equation}
\begin{equation}
{\overline {P_{t,i}^2 }  = \int_{P_{\min } }^{P_{\max } } {P_{t,i}^2
f_P \left( {P_{t,i} } \right)} dP_{t,i}}.
\end{equation}
\section{Estimation of the path loss between PU and SU}
\subsection{MMSE estimation of $L_{ps}$}
From Equation $(5)$, we can know that the inversion of path loss
between PU and SU $x$ and the SU received power vector ${\bf{P}}_r$
satisfy a linear relationship. Therefore, SU can estimate the
inversion of path loss between PU and SU $x$ by using linear MMSE
method \cite{c8} as
\begin{equation}
{\hat x = {\bf{R}}_{x{\bf{P}}_r } {\bf{R}}_{{\bf{P}}_r {\bf{P}}_r
}^{ - 1} {\bf{P}}_r},
\end{equation}
where  row vector ${\bf{R}}_{x{\bf{P}}_r }$ is the cross-correlation
between $x$ and ${\bf{P}}_r$. Its $i$-th element is
\begin{equation}
{{\bf{R}}_{x{\bf{P}}_r } (i) = E\left[ {xP_{r,i} } \right] = \bar
P_{t,i} \overline {x^2 }  + \sigma ^2 \bar x},
\end{equation}
where $\bar x$ and $\overline {x^2 }$ represent the mean value and
mean square value of $x$ respectively. The calculation of $\bar x$
and $\overline {x^2 }$ will be introduced in the next subsection.
${\bf{R}}_{{\bf{P}}_r {\bf{P}}_r }$ is the auto-correlation matrix
of ${\bf{P}}_r$. Its $i$-th diagonal element is
\begin{equation}
{\begin{array}{l}
 {\bf{R}}_{{\bf{P}}_r {\bf{P}}_r } \left( {i,i} \right) = E\left[ {P_{t,i}^2 } \right] \\
 \;\;\;\;\;\;\;\;\;\;\;\;\; = \overline {P_{t,i}^2 }  \cdot \overline {\left| {h_{ps,i} } \right|^4 }  \cdot \overline {x^2 }  + 2\bar P_{t,i}  \cdot \overline {\left| {h_{ps,i} } \right|^2 }  \cdot \bar x \cdot \sigma ^2  + 3\sigma ^4  \\
 \;\;\;\;\;\;\;\;\;\;\;\;\; = 2 \cdot \overline {P_{t,i}^2 }  \cdot \overline {x^2 }  + 2\bar P_{t,i}  \cdot \bar x\sigma ^2  + 3\sigma ^4  \\
 \end{array}},
\end{equation}
and its non-diagonal element $(i_1  \ne i_2)$
\begin{equation}
{\begin{array}{l}
 {\bf{R}}_{{\bf{P}}_r {\bf{P}}_r } \left( {i_1 ,i_2 } \right) = E\left[ {P_{r,i_1 } P_{r,i_2 } } \right] \\
 \;\;\;\;\;\;\;\;\;\;\;\;\;\;\;\;\; = \bar P_{t,i}^2 \overline {x^2 }  + \bar P_{t,i} \bar x\sigma ^2  + \bar P_{t,i} \bar x\sigma ^2  + \sigma ^4  \\
 \end{array}}.
\end{equation}
After getting the estimate of the inversion, the path loss from PU
to SU can be obtained as
\begin{equation}
{\hat L_{ps}  = \frac{1}{{\hat x}}}.
\end{equation}
The estimate of path loss from SU to PU $\hat L_{sp}$ can be
obtained from $\hat L_{ps}$ according to the adopted duplex mode, as
stated in the end of Section III-A.

\subsection{Calculation of the mean and mean square values of $x$}
When $\hat L_{pb}$ and propagation loss model are known, the
location of PU can be confined on a circle whose center is BS and
radius is $r_0  = {{(\hat L_{pb} } \mathord{\left/
 {\vphantom {{(\hat L_{pb} } {L_0 }}} \right.
 \kern-\nulldelimiterspace} {L_0 }})^{{1 \mathord{\left/
 {\vphantom {1 \alpha }} \right.
 \kern-\nulldelimiterspace} \alpha }}$. Suppose SU knows the distance between itself and BS $r_1$,
 which can be estimated by using the pilot signal broadcast by BS. Then, the distance between PU and
 SU is $D = \sqrt {r_0^2  + r_1^2  - 2r_0 r_1 \cos \theta }$, where
 $\theta$ is the angle between SU and PU, as shown in Fig. 2. Thus, the mean and mean square values of $x$ can be calculated as
 \begin{equation}
{\bar x = \int_0^{2\pi } {\frac{1}{{2\pi L_0 D^\alpha  }}} d\theta},
\end{equation}
\begin{equation}
{\overline {x^2 }  = \int_0^{2\pi } {\frac{1}{{2\pi (L_0 D^\alpha
)^2 }}d\theta }}.
\end{equation}

\begin{figure}[!t]
\centering
\includegraphics[width=0.3\textwidth]{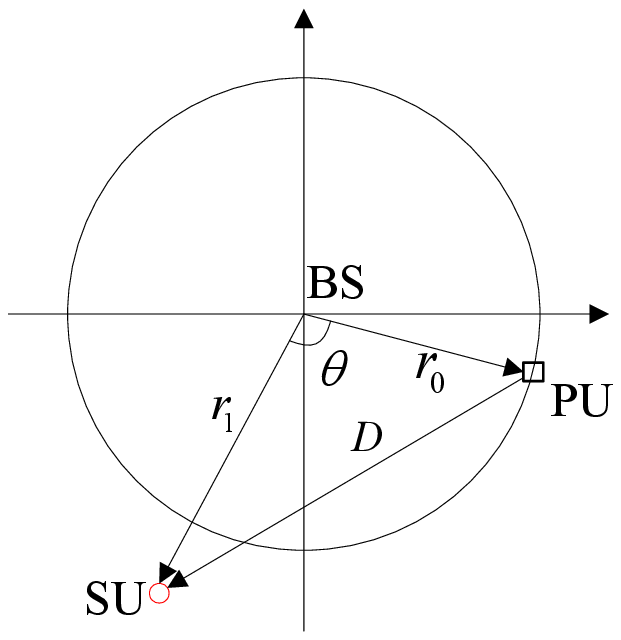}
\caption{Illustration of the position of PU and SU}
\label{PU_circle}
\end{figure}

\subsection{Estimation procedure}
The procedure of estimating the path loss between PU and SU $\hat
L_{ps}$ is summarized as follows.
\begin{enumerate}
\item SU demodulates and records the AMC mode sequence number
assigned by BS in the uplink and downlink $m_{u,i}$ and $m_{d,i}$ in
the past $I$ time instants;
\item By exploiting downlink AMC mode sequence number vector
${\bf{m}}_d$, SU estimates the downlink path loss between BS and PU
$\hat L_{bp}$ according to Equations $(10)$ and $(11)$;
\item By exploiting the estimated $\hat L_{bp}$ and uplink AMC mode sequence
number $m_{u,i}$, SU obtains the mean value $\bar P_{t,i}$ and mean
square value $\overline {P_{t,i}^2 }$ of $P_{t,i}$ according to
Equations $(13)$ and $(14)$, and the mean value $\bar x$ and mean
square value $\overline {x^2 }$ of $x$ according to Equations $(20)$
and $(21)$;
\item SU calculates the cross-correlation vector
${\bf{R}}_{x{\bf{P}}_r }$ between $x$ and ${\bf{P}}_r$ from Equation
$(16)$, and the auto-correlation matrix ${\bf{R}}_{{\bf{P}}_r
{\bf{P}}_r }$ of ${\bf{P}}_r$ from Equations $(17)$ and $(18)$;
\item According to the received power vector ${\bf{P}}_r$ in the past
$I$ time instants, SU first estimates the inversion of path loss
from PU to itself according to Equation $(15)$, and then obtains the
estimate of the path loss by using Equation $(19)$.
\end{enumerate}
\section{Simulation results}
Simulations are carried out to evaluate the performance of the
proposed path loss estimation method. Simulation parameters are set
as follows \cite{c9}. The radius of Macrocell is $500$ m, and BS is
deployed at the center of Macrocell. The minimum distance between PU
and BS is $35$ m. In downlink, BS transmits with a fixed power $P_0$
which ensures the average received SINR at the cell fringe to be $12
dB$. The propagation loss model is $L(d) = 15.3 + 37.6\log _{10} d$,
where $d$ represents the distance between two locations (unit: m).
According to the current channel status, BS assigns the AMC modes
which can satisfy the BER requirement and maximize the transmission
rate for transmission signals in downlink and uplink. The adopted
AMC modes are the same as those specified in IEEE 802.16e standard.
In uplink, for simulation convenience, the effect of AMC is not
considered. We assume that PU transmits with a variable power which
ensures the received SINR at any instant to be $15 dB$ exactly. The
number of considered time instants is $I=200$. Estimation error is
defined as the expectation of the absolute difference between
estimated value and real value, that is, $Error = E\left\{
{|real\;value - estimated\;value|} \right\}$, where the real and
estimated values are both expressed in $dB$.

The performance of the method to estimate the path loss between BS
and PU depends only on the distance between them. Fig. 3 shows the
relation between the estimation error and the distance. We can see
that the estimation error keeps within $1 dB$ when PU is far away
from BS (i.e., larger than $150$ m), while it becomes greater when
PU is closer to BS. This is because the distribution characteristics
of AMC mode changes obviously as the distance changes if PU is far
away from BS. However, when PU is close to BS, the quality of
received signal is always good, which makes BS always assign the AMC
modes with highest order for its transmission signals. Therefore, in
this case, it is very hard to estimate the path loss accurately
according to the distribution characteristics of AMC mode.

\begin{figure}[!t]
\centering
\includegraphics[width=0.5\textwidth]{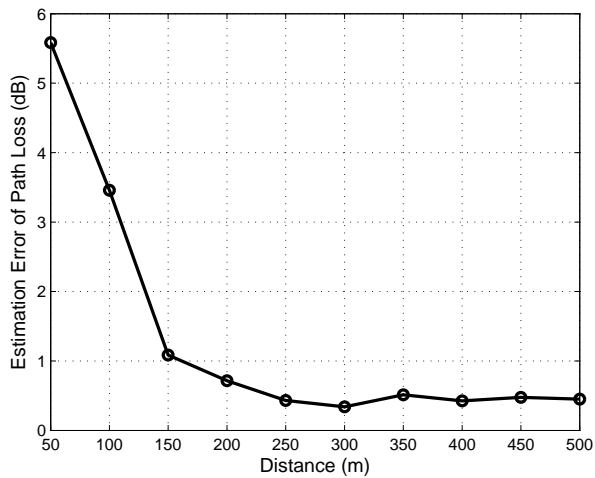}
\caption{Estimation error vs. the distance between PU and BS}
\label{BS_PU path loss}
\end{figure}

\begin{figure}[!t]
\centering
\includegraphics[width=0.35\textwidth]{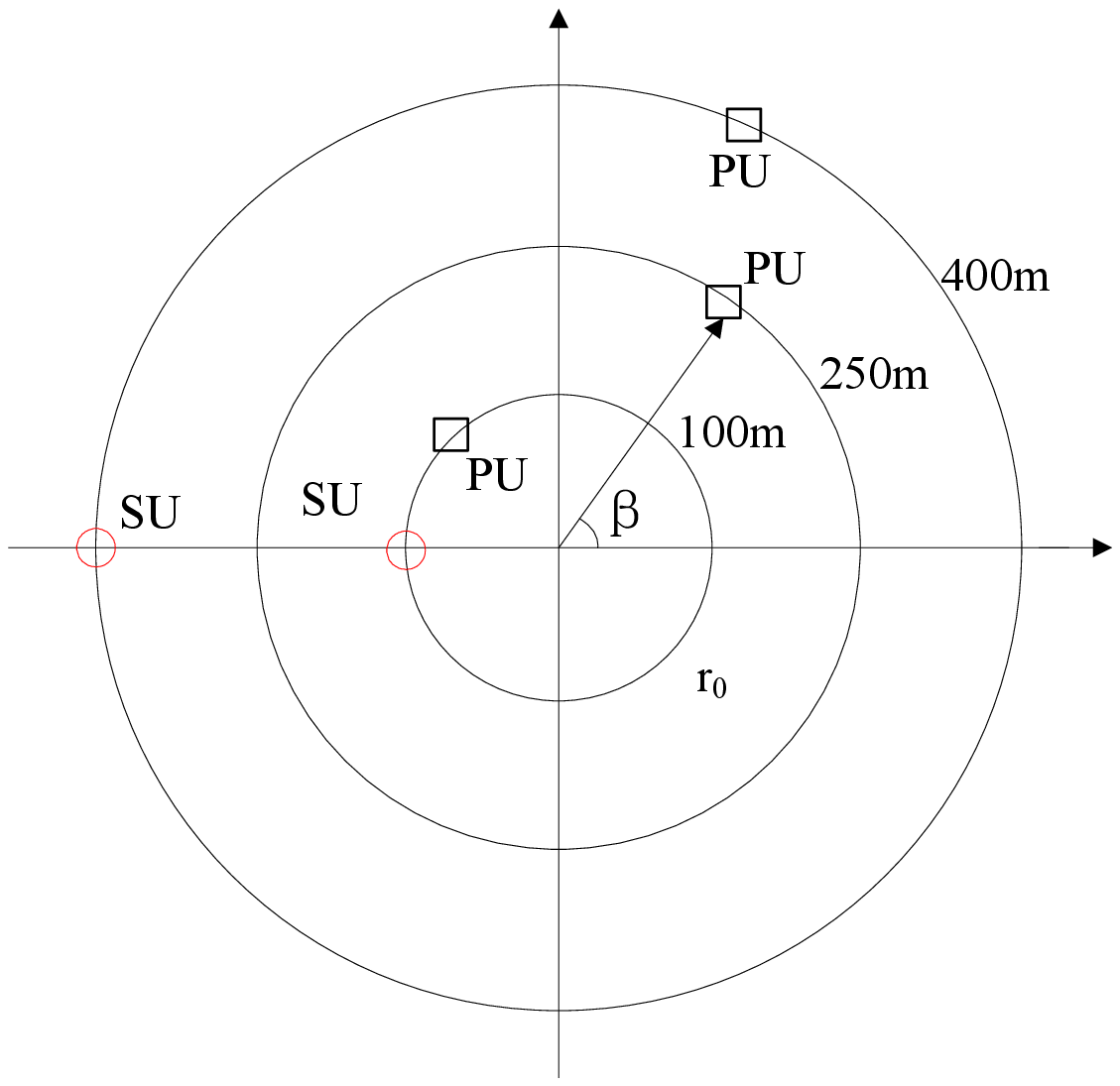}
\caption{Locations of SU, PU and BS} \label{position relation}
\end{figure}

\begin{figure}[!t]
\centering
\includegraphics[width=0.5\textwidth]{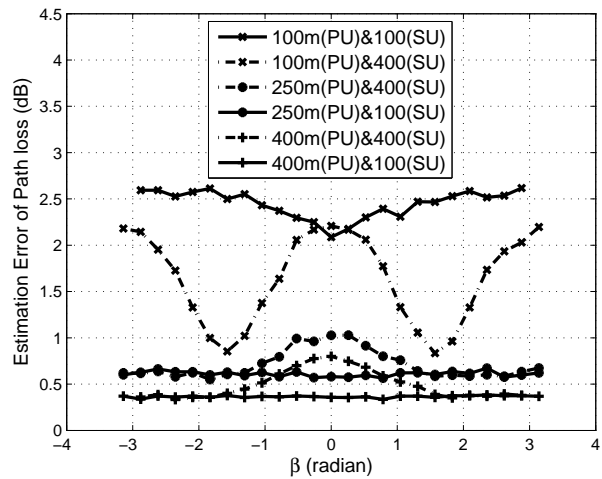}
\caption{Estimation error vs. PU location $(r_1=100m, 400m)$}
\label{PU_SU path loss}
\end{figure}

The performance of the method to estimate the path loss between PU
and SU depends on the locations of SU, PU and BS. The locations in
this simulation are shown in Fig. 4, where SU (marked as circle) is
placed at $(-100, 0)$ and $(-400, 0)$, which stand for the cases of
SU close to  and far from BS, respectively. PU (marked as square) is
confined at one of the three circles with radius $r_0$ equal to
$100$ m, $250$ m and $400$ m, respectively. For each location of SU,
the performance of the path loss estimation method is evaluated when
PU moves along each of the three circles.

Fig. 5 shows the relation of the estimation error of the path loss
between PU and SU versus the locations of PU when SU is located at
different places. It can be seen that the estimation error is always
below $2.7$ dB no matter where SU and PU are located, which
indicates that our proposed path loss estimation method is very
effective and applicable. It is also seen that the estimation error
is always smaller when PU is far away from BS ($r_0=250$ m and $400$
m) than that when PU is close to BS ($r_0=100$ m).  Two factors
contribute to this. First, when PU is far away from BS, the
estimation error of the path loss between BS and PU in the first
step is smaller (see Fig. 3). So the performance of the method to
estimate the path loss between PU and SU in the second step is
better. Second, the PU transmission power is large when PU is far
away from BS. So the SU received power is large correspondingly,
which results in high received SINR and low estimation error.
\section{Conclusions}
This paper has proposed a novel non-cooperative path loss estimation
method. In this method, SU independently derives the PDF and
statistics of PU transmission power by exploiting the channel
statics and broadcasted AMC mode information, and estimates the path
loss between itself and PU by exploiting the relation among
transmission power, received power and path loss. This method does
not need any information exchange between SU and PU and does not
make any impact on the communications in the macrocell. For the
two-tier network whose primary tier is WiMAX or LTE, this method is
compatible with the primary tier network completely and does not
require any modification to the existing standard, which increases
the applicability of the proposed method in real systems
significantly. Simulation results show that our proposed method can
estimate the path loss between PU and SU effectively no matter where
SU and PU are located.

The proposed non-cooperative path loss estimation method is also
applicable to other geography-overlapped and spectrum-shared
two-tier networks, such as a traditional network co-existing with
cognitive networks, etc.

Future work includes the research on the relationship between the
performance and the number of observed samples I.

\section*{Acknowledgement}
This work was jointly supported by National Basic Research
Development Program of China (973 Program, No. 2009CB320405), Nature
Science Foundation of China (60972058) and Motorola (China)
Technologies Ltd. The discussion from Dr. Hai Jiang at the
University of Alberta is appreciated.

\end{document}